\documentclass[pra,twocolumn,amsmath,amssymb,floatfix,reprint]{revtex4}

\usepackage{graphicx}
\usepackage{color}
\usepackage{pst-node}
\usepackage{bbold}

\def \cL{{\cal{L}}}

\def \br{{\bf r}}

\def \bq{{\bf q}}
\def \bk{{\bf k}}
\def \bp{{\bf p}}

\def \b1{{\bf 1}}

\def \cL{{\mathcal L}}

\newcommand{\bea}{\begin{eqnarray}}

\newcommand{\eea}{\end{eqnarray}}

\newcommand{\beq}{\begin{equation}}

\newcommand{\eeq}{\end{equation}}

\def \br{{\bf r}}

\def \bq{{\bf q}}
\def \bk{{\bf k}}
\def \bp{{\bf p}}

\def \bx{{\bf x}}

\def \be{{\bf e}}

\def \b0{{\bf 0}}

\newcommand{\unit}[1]{\,\mathrm{#1}}
\newcommand{\no}{\nonumber\\}

\newcommand{\ket}[1]{\left|#1\right>}

\newcommand{\expect}[1]{\langle #1 \rangle}
\newcommand{\abs}[1]{\left| #1 \right|}

\usepackage{ulem}

\newlength\figurewidth
\newlength\figurefullwidth
\begin{document}

\title{Squeezed-field path-integral description of second sound in Bose-Einstein condensates}

\begin{abstract}
We propose a generalization of the Feynman path integral using squeezed coherent states. We apply this approach to the dynamics of Bose-Einstein condensates, which gives an effective low energy description that contains both a coherent field and a squeezing field. We derive the classical trajectory of this action, which constitutes a generalization of the Gross-Pitaevskii equation, at linear order. We derive the low energy excitations, which provides a description of second sound in weakly interacting condensates as a squeezing oscillation of the order parameter. This interpretation is also supported by  a comparison to a numerical c-field method. 
\end{abstract}  

\author{Ilias M. H. Seifie, Vijay Pal Singh, and L.~Mathey}

\affiliation{
\mbox{Zentrum f\"ur Optische Quantentechnologien, 
Universit\"at Hamburg, 22761 Hamburg, Germany}\\
\mbox{Institut f\"ur Laserphysik, Universit\"at Hamburg, 22761 Hamburg, Germany}\\
\mbox{The Hamburg Center for Ultrafast Imaging, Luruper Chaussee 149, Hamburg 22761, Germany}
}

\date{\today}

\maketitle

\section{Introduction}

The Feynman path integral has been one of the most fruitful concepts of theoretical physics \cite{feynman_original}. 
It provides an alternative view of quantum mechanics by formulating it as a sum of paths.
Each of these paths is weighted by a phase given by the classical action, and therefore it recovers the Lagrangian method in the context of quantum mechanics, see also \cite{dirac}. Path-integral formulations have been applied to both dynamic and thermodynamic quantities, and to classical and quantum systems and processes.   Numerous analytical and numerical methods have been developed, see e.g. \cite{feynmanhibbs,schulman,kleinert}.

\begin{figure}[t]
\includegraphics[width=1.0\figurewidth]{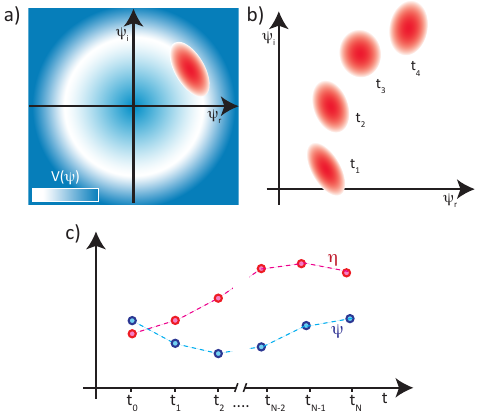}
\caption{%
 (a)  Illustration of the ordered state of a weakly interacting condensate, described by $|\phi|^{4}$ theory. The bosons condense in the minimum of the Mexican hat potential $V(\psi) = - \mu |\psi|^{4} + g |\psi|^{4}/2$, with $\mu, g>0$. The distribution, shown in red, is squeezed. Panel (b) shows a sketch of a single path in the path integral in which both the expectation value and the squeezing of the distribution vary in time. Panel (c) shows schematically the corresponding fields $\psi(t)$ and $\eta(t)$. 
}
\label{fig:cover}
\end{figure}

Further down, we present a generalization of the Feynman path integral, and exemplify this approach by
 applying it to complex $|\phi|^{4}$ theory. This is one of the most quintessential field theories of condensed matter, and is naturally realized in Bose-Einstein condensates of ultracold atoms. The dynamics of condensates continues to be an intriguing and subtle field of research. Phenomena such as superfluidity \cite{raman, desbuquois, moritz} and second sound of condensates \cite{grimm, dalibard, straten_meppelink} continue to pose questions, and are under recent and current investigation in ultracold atom systems. 
   The phenomenon and terminology of second sound was established for helium-II, and successfully described within a hydrodynamic two-fluid approach \cite{landau}. If the interactions in ultracold atom condensates are sufficiently strong to ensure local equilibration and therefore hydrodynamic dynamics, this approach is equally successful \cite{straten,nikuni_griffin}. However, due to the tunability of interactions of ultracold atoms, and the availability of bosonic atoms with very weak interactions, a non-hydrodynamic regime of dynamics can be reached. Here, even the terminology of first and second sound, inherited from the studies of helium-II, might have to be adjusted.  
  Theoretical studies on second sound in atomic condensates have been reported in \cite{pitaevskistringari, dmytruk, wetterich, ota, verney}.

  \begin{figure*}[t]
\includegraphics[width=2.08\figurewidth]{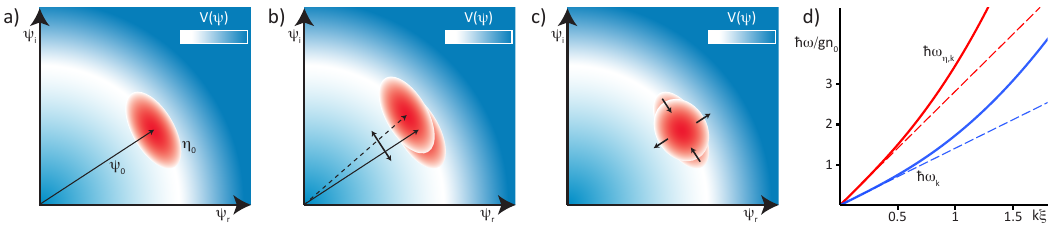}
\caption{%
(a) Illustration of the ground state, cf. Fig. \ref{fig:cover} (a), and its representation as an equilibrium state of the coherent field $\psi_{0}$ and the squeezing field $\eta_{0}$. Panel (b) shows the coherent field eigenmodes, which correspond to the Bogolyubov modes, and panel (c) shows the squeezing field eigenmodes, which are breathing modes around the equilibrium state. In panel (d), we show the dispersion of the squeezing field $\hbar \omega_{\bk,\eta}$ as a red continuous line, its low energy approximation $\hbar c_{2} |\bk|$ as a red, dashed line, the Bogolyubov mode $\hbar \omega_{\bk}$ as a blue line, and its low energy approximation $\hbar c_{1} |\bk|$ as a blue, dashed line. These are shown in units of the mean-field energy $g n_{0}$, and as a function of $k \xi$, where $\xi$ is the healing length $\xi = \hbar/\sqrt{2 m g n_0}$. 
}
\label{fig:modes}
\end{figure*}
 
In this paper, we generalize the path integral by extending the utilized set of states to squeezed coherent states. We exemplify this approach for $|\phi|^{4}$ theory, which results in a description of the weakly interacting Bose gas that explicitly captures the squeezed nature of its ordered state. An intuitive motivation is sketched in Fig. \ref{fig:cover} (a), which depicts the symmetry broken state of $|\phi|^{4}$ theory. It has an anisotropic distribution around its expectation value. To include this feature explicitly in the path integral, we utilize two-mode squeezed  coherent  states, instead of the commonly used coherent states. Therefore, the squeezed distribution of the equilibrium state is included in a single path. The resulting action that appears in the weight function for each path contains not only the complex field that describes coherent states, but an additional complex field that describes the squeezing of this field, see Fig. \ref{fig:cover} (c). We refer to these as the coherent and the squeezing field. A single path can be visualized as in Fig. \ref{fig:cover} (b). During the time evolution not only the expectation value of the distribution varies but also the quadratures around it. This visualizes that the information in a single path of this path integral, such as the classical path, contains information about higher order fluctuations than the regular coherent state path integral.

We note that the Bogolyubov approximation of the weakly interacting Bose gas uses two mode squeezing of the momenta $\bk$ and $-\bk$, in particular
\bea\label{eq:bogo}
b_{\bk} &=& u_{\bk}(\eta_{\bk}) \beta_{\bk} + v_{\bk}(\eta_{\bk}) \beta_{-\bk}^{\dagger}\, ,
\eea
where $b_{\bk}$ are the boson operators, and $u_{\bk}(\eta_{\bk})$ and $v_{\bk}(\eta_{\bk})$ are the Bogolyubov parameters, which both depend on a squeezing parameter $\eta_{\bk}$. However, the parameters $\eta_{\bk}$ are constant in the Bogolyubov approximation. They are chosen as $\eta_{\bk} = \eta_{\bk}^{0}$ to diagonalize the Hamiltonian, resulting in the Bogolyubov modes and their dispersion. 
 The equilibrium state characterized by these static squeezing parameters and the ground state condensate amplitude is visualized in    Fig. \ref{fig:modes} (a). 
In the path integral that we propose here, the squeezing parameters $\eta_{\bk}$ are allowed to evolve in time, and are itself a dynamical field. We derive the equations of motion for the classical path of the resulting Lagrange density, which generalizes the Gross Pitaevskii equation at linear order. We diagonalize the equations of motion, and derive the eigenmodes. The eigenmodes of the coherent field are the  Bogolyubov modes which are visualized in Fig. \ref{fig:modes} (b). In addition, we obtain the eigenmodes of the squeezing field, which are breathing modes of the equilibrium state, see Fig. \ref{fig:modes} (c), which we identify as second sound in the weakly interacting regime. 
 For asymptotically weak interactions, we demonstrate that the ratio of the Bogolyubov velocity $c_{1}$ and the second sound velocity $c_{2}$ is $c_{2}/c_{1} = 2$. 
  With increasing interaction strength, the magnitude of $c_{2}$ is reduced rapidly to values below $c_{1}$, as the system enters the hydrodynamic regime.  
   
   \begin{figure*}[t]
\includegraphics[width=2.1\figurewidth]{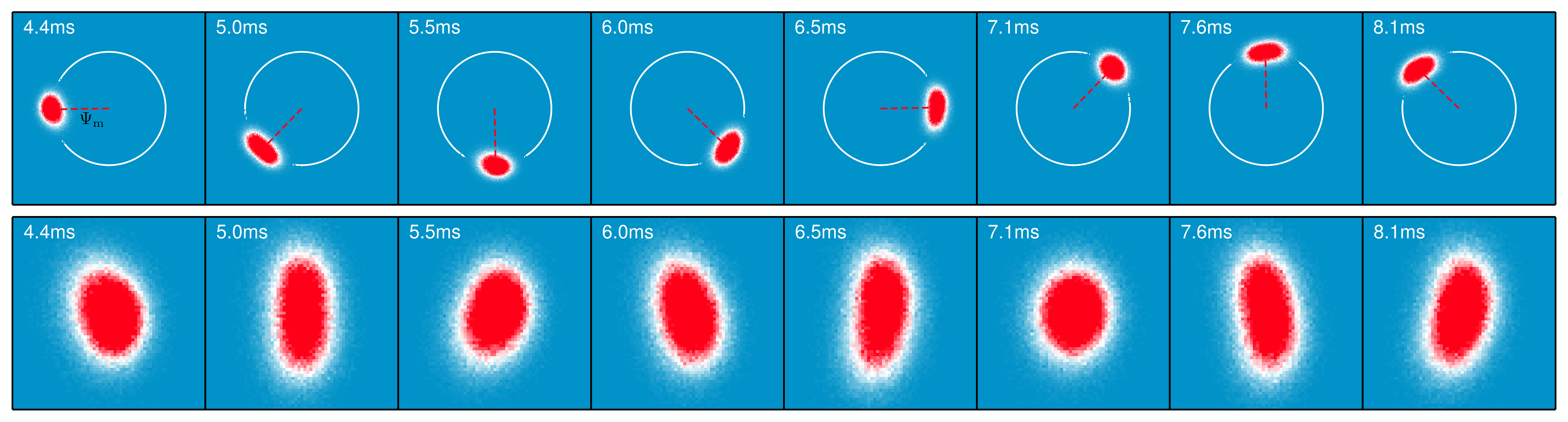}
\caption{%
Histogram of the field $\psi_{i}(t)$ at times $t_{j}$, following a momentum kick at time $t=0$, as given in Eq. \eqref{kick}. For the full time evolution see Ref \cite{SuppMat_animation}. In the upper row, we depict the full distribution, in the lower row, we remove the rotational motion. The white circle has the radius $\sqrt{n_{0}}$, where $n_{0}$ is the numerically determined condensate fraction. The dynamics consist of a rotation of the cloud around the origin, as well as a breathing motion, which we capture within the squeezed field approach. 
}
\label{fig:histogram}
\end{figure*}

\section{Squeezed field path integral}
   
We consider the Hamiltonian
\bea
H & =& \sum_\bk \epsilon_\bk b_\bk^{\dag}b_\bk +\frac{g}{2V}\sum_{\bk,\bp,\bq}b_{\bk+\bq}^{\dag}b_{\bp-\bq}^{\dag}b_\bk b_\bp\, ,\label{Ham}
\eea
where $\epsilon_{\bk} = \hbar^{2}k^{2}/(2 m)$ is the dispersion, $m$ the atom mass, $g$ the interaction strength, and $V$  the volume.  
We derive the  propagator $G$ of this system, defined as 
\bea\label{eq:propagator}
i G(\Psi_b, t_b;\Psi_a, t_a)&=&\langle\Psi_b |e^{-i H (t_{b}-t_{a})/\hbar)}|\Psi_a\rangle.
\eea
$\Psi_{a}$ refers to the initial state at time $t_{a}$, $\Psi_{b}$ to the final  state at time $t_{b}$, with $t_{b} > t_{a}$. The prefactor $i$ of the propagator is introduced following standard convention. We use  states of the form
\bea\label{eq:squeezecoherent_state}
\ket{\Psi}&=& \left[ {\textstyle \prod{\vphantom\prod}'}_{\bk\neq 0} S_\bk D_\bk D_{-\bk} \right]D_{\b0}\ket{0},
\eea 
for $|\Psi_{a/b}\rangle$, with  coherent state operators $D_\bk \equiv \exp(\psi_\bk b_\bk^{\dag}-\psi_\bk^*b_\bk)$ for each momentum mode $\bk$, and two-mode squeezing operators $S_\bk \equiv \exp(\eta _\bk b_\bk^{\dag}b_{-\bk}^{\dag} - \eta^*_\bk b_\bk b_{-\bk})$ for each pair of momentum modes $\bk$ and $-\bk$, with $\bk\neq 0$. The state \eqref{eq:squeezecoherent_state} is  a product of a coherent state with $\psi_{\b0}= \sqrt{N_{0}}$ for the $\bk = 0$ mode, which is fixed throughout the derivation, and a product of coherent squeezed states for all other momentum states. The product operation ${\textstyle  \prod{\vphantom\prod}'}_{\bk\neq 0}$  refers to all momentum states, but excludes double counting \cite{footnote}. For fixed $\psi_{\b0}$, these states resolve the identity in momentum space, excluding $\bk = 0$, i.e., $\mathbb{1}_{\bk\neq 0}\otimes |\Psi_{0} \rangle \langle \Psi_{0}|  = \int  d \Psi  |\Psi \rangle \langle \Psi|$. The integration measure is $d \Psi = {\textstyle \prod{\vphantom\prod}'}_{\bk\neq 0} d^2\eta_{\bk} \prod_{\bk\neq 0} d^2\psi_{\bk}/(C^{(N_{\bk}-1)/2}\pi^{2 (N_{\bk}-1)})$, where $N_{\bk}$ is the number of momentum modes, and $C$ is the area of the complex plane. We split the time interval into $N$ intervals of length $\Delta t\equiv (t_b-t_a)/N$, and introduce the resolution of the identity $N-1$ times, 
\bea\label{eq:propagatorsplit}
i G(\Psi_b, t_b;\Psi_a, t_a)&=&
\int  \mathcal{D}  \Psi \prod_{j=1}^{N}
\langle\Psi_j |e^{-i H \Delta t/\hbar)}|\Psi_{j-1}\rangle\, ,\quad
\eea
with $ \mathcal{D}  \Psi = \prod_{j=1}^{j=N-1} d \Psi_{j}$, $t_{N}\equiv t_{b}$ and $t_{0}\equiv t_{a}$. This is schematically shown in Fig. \ref{fig:cover} (c). To take the continuum limit $\Delta t \rightarrow 0$, we approximate each factor 
$\langle\Psi_j |e^{-i H \Delta t/\hbar)}|\Psi_{j-1}\rangle \approx  e^{- i \langle\Psi_j | H|\Psi_{j}\rangle \Delta t/\hbar}  \langle\Psi_j |\Psi_{j-1}\rangle$. The overlap $\expect{\Psi_j |\Psi_{j-1}}$ and its expansion to first order in $\Delta\Psi_j\equiv\Psi_j - \Psi_{j-1}$ is given in Eqs. \eqref{eq:resultscalarproduct}$-$\eqref{eq:prefactor_a}. 
In the continuum limit, the product of overlaps approaches
 \bea\label{eq:overlap_continuum}
 \prod_{j=1}^{N}  \langle\Psi_j |\Psi_{j-1}\rangle
  & \rightarrow & \exp\Big[ \int_{t_{a}}^{t_{b}} dt \Big(\sum_{\bk \neq 0} \frac{\psi_{\bk} \partial_{t}\psi_{\bk}^{*} -\psi^{*}_{\bk} \partial_{t}\psi_{\bk}}{2}\nonumber\\
&&  + \sum_{\bk\neq 0}{\vphantom\sum}'(a_{\bk} \partial_{t}\eta_{\bk}^{*} - a^{*}_{\bk} \partial_{t}\eta_{\bk}) \Big)\Big]\, .
 \eea
The continuum limit transforms the discrete sequence of parameters $\psi_{\bk, j}$ and $\eta_{\bk, j}$ into $\psi_{\bk}(t)$ and $\eta_{\bk}(t)$. The function  $a_{\bk}=a_{\bk}(\psi_{\bk}, \psi_{-\bk}, \eta_{\bk})$ that couples to the time derivative of $\eta_{\bk}(t)$ is given in Eq. \eqref{eq:prefactor_a}.

To evaluate $\langle\Psi_j | H|\Psi_{j}\rangle$, we first expand the interaction term to second order in the operators $b_{\bk}$ with $\bk \neq 0$, then evaluate the expectation value  of the state $|\Psi_{j}\rangle$, and expand to second order in the coherent state amplitudes $\psi_{\bk}$ around $\psi^0_{\bk}=0$. We expand  $\eta_\bk$ around its equilibrium value to second order, i.e.,  $\eta_{\bk} = \eta_{\bk}^{0} + \tilde{\eta}_{\bk}$. The equilibrium value is $\eta_{\bk}^{0} = - \ln(1+2 g n_0/\epsilon_{\bk})/4$.  This value solves $g n_0 (u_{\bk, 0}^{2}+|v_{\bk,0}|^{2})/2 + (\epsilon_{\bk} + g n_0)u_{\bk,0} v_{\bk,0} =0$,  where $u_{\bk,0}= u_{\bk}(\eta_{\bk}^{0})$ and $v_{\bk,0}= v_{\bk}(\eta_{\bk}^{0})$, and diagonalizes the Bogolyubov modes. These two expansions give
$\langle\Psi_j | H|\Psi_{j}\rangle = H_{\psi} + H_{\eta}$, with $H_{\psi} =  \sum_{\bk\neq 0} \hbar \omega_{\bk} |\psi_{\bk}|^{2}$, where $\hbar\omega_{\bk} = \sqrt{\epsilon_{\bk}(\epsilon_{\bk} + 2 g n_{0})}$ is the standard Bogolyubov dispersion and $n_{0}$ is the condensate density. For the squeezing field, we have $H_{\eta} = \sum_{\bk\neq 0}' (E_{\bk, r} \tilde{\eta}_{\bk, r}^{2} + E_{\bk, i} \tilde{\eta}_{\bk,i}^{2})$,  as derived in the App. \ref{appA3}, where we also give the full expressions for $E_{\bk, r}$ and $E_{\bk, i}$.  The fields $\tilde{\eta}_{\bk, r/i}$ are the real/imaginary part of $\tilde{\eta}_{\bk}$. 
Combining these results, the propagator takes the form
$i G(\Psi_b, t_b;\Psi_a, t_a)=\int  \mathcal{D}  \Psi \exp( i S/\hbar)$, with the action $S = \int_{t_{a}}^{t_{b}} dt \cL$. 
The Lagrangian is $\cL = \cL_{\psi} + \cL_{\eta}$ with
\bea\label{eq:lagrangians}
\cL_{\psi} &=& \sum_{\bk \neq 0} \frac{i\hbar}{2}(\psi^{*}_{\bk} \partial_{t}\psi_{\bk}-\psi_{\bk} \partial_{t}\psi_{\bk}^{*} ) - \hbar \omega_{k}|\psi_{k}|^{2}\\
\cL_{\eta} &=& \sum_{\bk \neq 0}{\vphantom{\sum}}' i\hbar(a^{*}_{\bk} \partial_{t}\tilde{\eta}_{\bk}-a_{\bk} \partial_{t}\tilde{\eta}_{\bk}^{*}   ) - E_{\bk, r} \tilde{\eta}_{\bk, r}^{2} - E_{\bk, i} \tilde{\eta}_{\bk,i}^{2}.\quad\,\,
\eea
The coherent field has the dispersion $\hbar \omega_{\bk}$, recovering the standard Bogolyubov result. 
 To diagonalize $\cL_{\eta}$, we expand $a_{\bk}$ to first order, i.e. $a_{\bk} \approx  \textnormal{const.}  + a_{\bk, 1} \tilde{\eta}_{\bk} + \bar{a}_{\bk, 1} \tilde{\eta}^{*}_{\bk}$, where $a_{\bk, 1}$ and $\bar{a}_{\bk, 1}$ are real-valued expansion coefficients independent of $\tilde{\eta}_{\bk}$, given in the App. \ref{appA3}. Using the substitution $\tilde{\eta}_{\bk, r} = (E_{\bk, i}/(4 E_{\bk, r}))^{1/4} (\xi_{\bk} + \xi_{\bk}^{*})$ and $\tilde{\eta}_{\bk, i} = - i(E_{\bk, r}/(4 E_{\bk, i}))^{1/4} (\xi_{\bk} - \xi_{\bk}^{*})$, we obtain  
 \bea
  \cL_{\eta} &=& \sum_{\bk \neq 0}{\vphantom\sum}' 4 a_{\bk, 1}\Big[\frac{i\hbar}{2}(\xi^{*}_{\bk} \partial_{t}\xi_{\bk}- \xi_{\bk} \partial_{t}\xi_{\bk}^{*}   ) - \hbar \omega_{\bk,\eta} |\xi_{\bk}|^{2} \Big],\qquad
\eea
where $\hbar \omega_{\bk,\eta} = \sqrt{E_{\bk,r} E_{\bk,i}}/(2 a_{\bk, 1})=2\omega_\bk$ is the dispersion of the squeezing mode. The low-frequency limit of this dispersion is  $ \hbar \omega_{\bk,\eta} \approx    \hbar c_{2} |\bk|$
 with  $c_{2} = 2 c_{1}$. These dispersions are shown in Fig. \ref{fig:modes} (d).

\section{Single-particle Green's function} 

 To elaborate on the physical consequences of these modes, we determine the single-particle Green's function    
    $i g_1(\bk,t_{2}, t_{1}) = \langle 0 | T(b_{\bk}(t_{2}) b^{\dagger}_{\bk}(t_{1})) | 0 \rangle$ at zero temperature, where $T$ is the time ordering operator.
        In the squeezed field path integral formalism this is
        $i g_1(\bk,t_{2}, t_{1}) = \langle \beta_{\bk}(t_{2}) \beta^{*}_{\bk}(t_{1}) \rangle$
        with $\beta_{\bk}(t) = u_{\bk}(\eta_{\bk}(t)) \psi_{\bk}(t) +  v_{\bk}(\eta_{\bk}(t)) \psi^{*}_{-\bk}(t)$.
        We expand $u_{\bk}(\eta_{\bk}(t))$ and $v_{\bk}(\eta_{\bk}(t))$ to first order around $\eta_{\bk}(t)= \eta_{\bk}^{0}$. 
        We use the Green's functions of $\psi_{\bk}$ and $\xi_{\bk}$, which are $\langle\psi_{\bk, \omega}^{*}\psi_{\bk, \omega}\rangle = i/(\omega-\omega_{\bk} + i \delta)$ and $\langle\xi_{\bk, \omega}^{*}\xi_{\bk, \omega}\rangle = i (4 a_{\bk, 1})^{-1}/(\omega-\omega_{\eta, \bk} + i \delta)$ in frequency space, and obtain
 \bea 
 g_1(\bk,\omega) & = & g_1^B(\bk,\omega) + \frac{g^{(+)}_{\bk}}{\omega-\omega^{+}_{\bk} + i \delta}
  - \frac{g^{(-)}_{\bk}}{\omega+\omega^{+}_{\bk} - i \delta},\qquad\label{G}
 \eea
  see App. \ref{appA4}. Here, $g_1^B(\bk,\omega)$ is the standard Bogolyubov Green's function, i.e. $g_1^B(\bk,\omega) = u^{2}_{\bk,0}/(\omega- \omega_{\bk} + i \delta) - v^{2}_{\bk,0}/(\omega+\omega_{ \bk} - i \delta)$, and $g^{(+)}_{\bk}$ and $g^{(-)}_{\bk}$ are given in 
    Eq. \eqref{eq:prefacs}.  
      The Green's function $g_1(\bk,\omega)$ displays 
 additional side peaks at $\pm\omega_\bk^+=\pm (\omega_{\bk} + \omega_{\eta,\bk})$, which are the modified response of the single-particle Green's function due to the squeezing mode.

\begin{figure}[t]
\includegraphics[width=00.9\figurewidth]{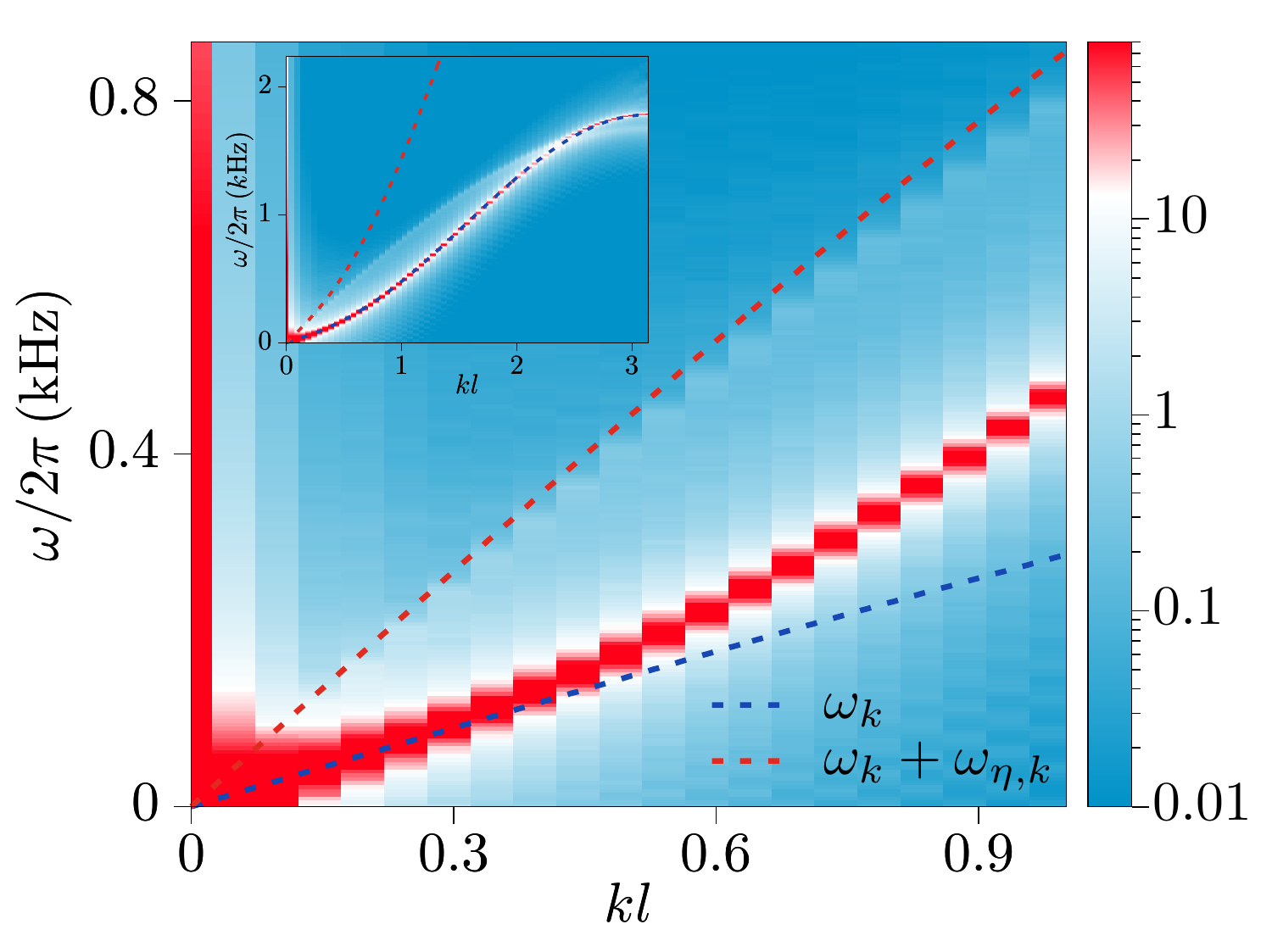}
\caption{%
  Single-particle correlation function $c_{1}(\bk, \omega)$ as a function of frequency $\omega$ and momentum $k_{x}$, for the interaction $U/J=0.05$.  The two excitation branches are compared to the Bogolyubov dispersion $\omega_{\bk}$, blue dashed line, and the side peak at $\omega_{\bk} + \omega_{\bk, \eta}$, red dashed line, motivated by the pole structure of Eq. \eqref{G}. The inset shows half of the Brillouin zone, the large figure the low-energy regime, where the approximations $\omega_{\bk}\approx c_{1}|\bk|$ and $\omega_{\bk,\eta}\approx c_{2}|\bk|$ are used. 
}
\label{fig:kinEn}
\end{figure}
 
\section{Comparison to c-field simulation} 
 
As a comparison, we consider a numerical implementation of the c-field method, see Refs. \cite{singh, mathey}.
We discretize space with discretion length $l$ and approximate the system, Eq. \eqref{Ham}, with a Hubbard model
\beq
H_{l} = -J \sum_{\langle i j\rangle} (b_{i}^{\dagger}b_{j} + h.c.) - \sum_{i} \mu  n_{i}  + \frac{U}{2} \sum_{i}   b_{i}^{\dagger}b_{i}^{\dagger}b_{i}b_{i}\, ,
\eeq
with $n_{i} = b_{i}^{\dagger}b_{i}$. 
The Hubbard parameters are related to the continuous space parameters via $J = \hbar^{2}/(2 m l^{2})$ and $U = g/l^{3}$. We choose a lattice with the dimensions $N_{x}\times N_{y}\times N_{z} = 128\times 32\times 32$. 
In the semi-classical approximation that we use here, we replace the bosonic operators $b_{i}$ with a complex field $\psi_{i}$. 
We choose $U/J =0.13$, and a chemical potential so that $\langle n_{i}\rangle \approx 7$. For a discretization length $l = 1 \mu$m and $^{7}$Li atoms, this corresponds to a real space density $\rho = 0.7\times 10^{13}$cm$^{-3}$, and a scattering length $a = 5.2$nm, which can be achieved by tuning near a zero crossing of the contact interaction, see \cite{Rem}. 
We initialize the state via Monte Carlo sampling at a temperature $T/J =1$, and propagate the equations of motion for $\psi_{i}(t)$.
At time $t_{k}$, which we define to be zero, we perform a momentum kick of the form
\bea
\psi_{i} &\rightarrow& \sqrt{1-A^{2}} \psi_{i}  + A \exp(i \bk \bx_{i}) \psi_{i}\, ,\label{kick}
\eea
with an amplitude $A=0.1$, and a momentum $\bk = (2\pi \times 12/N_{x})\be_{x}$, and integrate the subsequent time evolution. 
In Fig. \ref{fig:histogram}, we show the histogram of the field amplitudes $\psi_{i}$, depicted in the complex plane, with a binning size of $0.02\times 0.02$. The dynamical evolution that emerges after a few cycles consists of a rotation of the ensemble around the origin, which is depicted in the upper row. For the lower row, we determine the mean value of the distributions $\Psi_{m} = \frac{1}{N_l} \sum_i \psi_{i}$, where $N_l$ is the number of sites $N_l=N_xN_yN_z$, and determine its phase $\phi_{m}=\arctan(\Psi_{m, i}/\Psi_{m, r})$.
 In Fig. \ref{fig:histogram}, we connect $\Psi_{m}$ and the origin with a red, dashed line. 
 We rotate the distribution so that the center of mass of the ensemble is on the real, positive axis, and zoom in. We observe that in addition to the rotation of the ensemble, which corresponds to first sound, there is a breathing motion which corresponds to second sound.

As a second comparison, we determine the single-particle correlation function $c_{1}(\bk, \omega) = \langle  b^{\dagger}(\bk, \omega)  b(\bk, \omega)   \rangle$, where 
\beq
b(\bk, \omega) =\frac{1}{\sqrt{N_{l} T_{s}}} \sum_{i} \int dt e^{- i\bk \br_{i} - i \omega t}  b_{i}(t)\, .
\eeq
$T_{s}=227$ms is the sampling time for the numerical Fourier transform.
 In Fig. \ref{fig:kinEn}, we show  $c_{1}(\bk, \omega)$ for $\bk = k \be_{x}$, for a $^{7}$Li condensate with density $\rho = 0.6\times 10^{13}$cm$^{-3}$ and temperature $T/J = 4.5$. We observe two excitation branches in the numerical result, and compare them to the Bogolyubov dispersion $\omega_{\bk}$, and the side band $\omega_{\bk}^+$ that was found in Eq. (\ref{G}).
 We find good agreement in the low energy regime, which supports our squeezed field approach to the understanding of second sound in the weakly interacting regime. 
 We note that higher order terms of the Lagrangian will couple these modes and renormalize this weak coupling limit of the dispersions. This is most pronounced for the squeezing mode excitations with energies above the mean field energy, for which the breathing motion shown in Fig. \ref{fig:modes} (c) is reduced in energy compared to the linearly approximated potential around the mean value. This 
 and the thermal dependence of the dispersion will be discussed elsewhere. In  \cite{SuppMat_appB} we show the correlation function $c_{1}(\bk, \omega)$ for increasing interaction strength. The upper branch is rapidly renormalized below the Bogolyubov branch, indicating the emergence of the hydrodynamic regime. 

\section{Conclusions}
 In conclusion, we have developed a generalized path integral that utilizes squeezed coherent states, and have applied it to the weak coupling limit of Bose-Einstein condensates. 
 We have obtained the corresponding Lagrangian,  at linear order, which contains both the standard coherent field as well as an additional squeezing field. 
 We have derived the equations of motion, and determined the low-energy excitations of the condensed state.  One of the two excitation branches recovers the Bogolyubov modes, the other one provides an analytical estimate for the second sound dispersion in the weak coupling limit. 
  Furthermore, it provides an interpretation of the phenomenon of second sound as a squeezing oscillation of the order parameter. 
 We note that the method that we have presented here is of broad applicability. It is of conceptual importance, because the same system, described by the same Hamiltonian, gives different generalized Lagrangians in the path integral, depending on the set of states that is used. As result, higher order quantum fluctuations are captured in the corresponding classical path. 
 Furthermore, any analytical approach that is based on a path integral representation can be generalized in the way that we have presented here.  Finally, any numerical method that derives from a path integral representation can be generalized by extending the set of states of the path integral, for which we have paved the way in this paper.


\begin{acknowledgments}
We acknowledge support from the Deutsche Forschungsgemeinschaft through the SFB 925 and the Hamburg Centre for Ultrafast Imaging, and from the Landesexzellenzinitiative Hamburg, which is supported by the Joachim Herz Stiftung.

\end{acknowledgments}


\appendix

\section{Analytical calculations}

Generally, the path integral is constructed by discretizing the propagator in time. This leads to two central objects that determine the Lagrangian, which are the overlap of two states at succeeding times $\expect{\Psi_j | \Psi_{j-1}}$ and the expectation value of the Hamiltonian $\expect{\Psi_j | H | \Psi_{j}}$. In the continuum limit, we then obtain a classical continuous Lagrangian. In order to calculate the low energy dispersions of the coherent and squeezing field, we then expand our expressions around the equilibrium values of our fields and derive the corresponding low energy equations of motion. Below, we present our calculations step by step. 

\subsection{Overlap $\expect{\Psi_j | \Psi_{j-1}}$}\label{appA1}

We start our calculation with the overlap of the squeezed coherent state at succeeding times. The total state at time step $j$ reads
\bea\label{eq:squeezecoherent}
\ket{\Psi}&=& \left[ {\textstyle \prod'}_{\bk\neq 0} S_{\bk j} D_{\bk j} D_{-\bk j} \right]D_{\b0}\ket{0} .
\eea 
The above operators commute for different $\bk-$modes, so it is sufficient to consider a single factor, i.e., a single pair of modes $(\bk,-\bk)$. Taking then the product over $\bk$, will lead to a summation in the exponent of the exponential. We derive the squeezed coherent state in terms of a creation operator acting on the vacuum $\ket{0}$, based on the single mode calculation in Ref. \cite{vogel}. It is helpful to consider the coherent squeezed state, i.e., first squeezing the vacuum, then displacing it into the complex plane, instead of the squeezed coherent state. Therefore, we introduce into the state $\ket{\Psi}$ in Eq. \eqref{eq:squeezecoherent} the identity  $S_{\bk}S_{\bk}^{\dag}=\unit{1}$ and obtain $S_{\bk}D_{\bk}S_{\bk}^{\dag}\,\, S_{\bk}D_{-{\bk}}S_{\bk}^{\dag}\,\, S_{\bk}\ket{0}$ for a specific $\bk-$mode. The squeezed coherent state is thus equivalent to the coherent squeezed state, when we squeeze the displacement operators $D_{\pm\bk}\rightarrow S_{\bk}D_{\pm \bk}S_{\bk}^{\dag}$. According to Ref. \cite{vogel}, the creation and annihilation operators transform under squeezing in the following way
 \beq\label{eq:squeezetrafo}
S_{\bk}b_{\pm {\bk}}S^{\dag}_{\bk}=u_{\bk}b_{\pm {\bk}}-v_{\bk}b_{\mp {\bk}}^{\dag},
\eeq
where $u_{\bk}=\cosh{\left(\left|\eta_{\bk}\right|\right)}$ and $v_{\bk}=e^{i\phi_{\eta_{\bk}}}\sinh{\left(\left|\eta_{\bk}\right|\right)}$. We can pull the squeezing operators into the exponent of the displacement operator by using the unitarity property of the squeezing operators. This will lead us to the standard displacement operator, but with renormalized coherent parameters $\psi_{\bk}^{\prime}=u_{\bk}\psi_{\pm \bk}+v_{\bk}\psi_{\mp \bk}^*$, which inherit the two modes coupling from the squeezing operator. Next, we decompose our squeezing and displacement operators into their normal ordered forms
\bea
D_{\pm \bk}&=&e^{\psi_{\pm \bk}b^{\dag}_{\pm \bk}}e^{-\psi_{\pm \bk}^*b_{\pm \bk}}e^{-\frac{\left|\psi_{\pm \bk}\right|^2}{2}}\label{eq:normalordered_coh}\\
S_{\bk}\,\,\,&=&e^{\frac{v_{\bk}}{u_{\bk}}b^{\dag}_{\bk}b^{\dag}_{-\bk}}\left(\frac{1}{u_{\bk}}\right)^{\hat{n}_{\bk}+\hat{n}_{-\bk}+1}e^{-\frac{v_{\bk}^*}{u_{\bk}}b_{\bk}b_{-\bk}}\label{eq:normalordered_sque}.
\eea
After using the Baker-Cambell-Hausdorff-Formula $\exp{X}\exp{Y}=\exp{\sum_{m=0}^{\infty}\frac{1}{m!}[X,Y]_m}\exp{X}$, we find the desired representation for our squeezed state
\bea\label{eq:staterepresentation}
&&\ket{\eta_{\bk}\psi_{\bk}\psi_{-\bk}}=D(\psi_{\bk}^{\prime})D(\psi_{-\bk}^{\prime})S_{\bk}\ket{0}=\no
&&\frac{1}{u_{\bk}}\exp\Biggl(\frac{v_{\bk}}{u_{\bk}}(b_{\bk}^{\dag}-\psi_{\bk}^{*\prime})(b_{-\bk}^{\dag}-\psi_{-\bk}^{*\prime})\no
&&+\psi_{\bk}^{\prime}b_{\bk}^{\dag}+\psi_{-\bk}^{\prime}b_{-\bk}^{\dag}-\frac{1}{2}(\left|\psi_{\bk}^{\prime }\right|^2+\left|\psi_{-\bk}^{\prime}\right|^2)\Biggl)\ket{0}\label{eq:staterepresentation}.
\eea
By projecting state \eqref{eq:staterepresentation} onto a two mode coherent state $\ket{\alpha_{\bk}\alpha_{-\bk}}$ with the closure relation and making use of the definition of coherent states, we can write the $\bk-$th factor of our scalarproduct as a $c-$number valued Gaussian integral which can be calculated via the relation \cite{altlandsimons} 
\beq\label{eq:gaussianrelation}
\int dz\, dz^* e^{-zw^*z+u^*z+z^*v}=\frac{\pi}{w}e^{\frac{u^*v}{w}}.
\eeq
This procedure yields the following expression for the overlap: 
\bea\label{eq:resultscalarproduct}
&&\expect{\eta_{\bk j}\psi_{\bk j}\psi_{-\bk j}|\eta_{\bk j-1}\psi_{\bk j-1}\psi_{-\bk j-1}}=\no
&&\frac{1}{u_{\bk j}u_{\bk j-1}-v_{\bk j}^*v_{\bk j-1}}\exp\Biggl(\frac{1}{u_{\bk j}u_{\bk j-1}-v_{\bk j}^*v_{\bk j-1}}\Biggl\{\no
&&-\frac{(u_{\bk j-1}u_{\bk j}-v_{\bk j-1}v_{\bk j}^*)}{2}
\bigg(\abs{\psi_{\bk j-1}}^2+\abs{\psi_{-\bk j-1}}^2+\no
&&\abs{\psi_{\bk j}}^2+\abs{\psi_{-\bk j}}^2\bigg)+\psi_{\bk j-1}\psi_{\bk j}^*+\psi_{-\bk j-1}\psi_{-\bk j}^*\no
&&+(u_{\bk j-1}v_{\bk j}^*-v_{\bk j-1}^*u_{\bk j})\psi_{\bk j-1}\psi_{-\bk j-1}\no
&&+(v_{\bk j-1}u_{\bk j}-u_{\bk j-1}v_{\bk j})\psi_{\bk j}^*\psi_{-\bk j}^*\Biggl\}\Biggl).
\eea
In the following, we write this state in terms of $\Delta\psi_{\bk j}\equiv \Psi_j-\Psi_{j-1}$. To this end, all parameters are expanded up to first order at time step $j$ around their values at the previous step $j-1$: The coherent parameter is replaced by $\psi_{\bk j}=\psi_{\bk j-1}+\Delta\psi_{\bk j}$ and the two Bogolyubov parameters ($u_{\bk j},v_{\bk j}$) are expanded as functions of the real and imaginary part of $\eta_{\bk j}$. Since we consider the exponent, i.e.,  $\ln\{\text{Eq. } \eqref{eq:resultscalarproduct}\}$, the global prefactor takes the form $\ln\{1/x\}$ and the prefactor inside of the exponent the form $1/x$. Both are expanded around $x_0=1$ because of the bosonic relation $u_{\bk j}^2-\abs{v_{\bk j}}^2=1$. We are left with an expression that depends only on the differences $\Psi_j-\Psi_{j-1}$
\bea\label{eq:resultscalarproduct_diff}
&&\ln\{\expect{\eta_{\bk j}\psi_{\bk j}\psi_{-\bk j}|\eta_{\bk j-1}\psi_{\bk j-1}\psi_{-\bk j-1}}\}\approx\no 
&&\frac{1}{2}\Big(\psi_{\bk j}\Delta\psi_{\bk j}^*-\psi_{\bk j}^*\Delta\psi_{\bk j}+\no
&&\psi_{-\bk j}\Delta\psi_{-\bk j}^*-\psi_{-\bk j}^*\Delta\psi_{-\bk j}\Big)\no
&&+a_{\bk j}\Delta\eta_{\bk j}^*-a_{\bk j}^*\Delta\eta_{\bk j},
\eea
where the parameter $a_{\bk j}$ has the following expression
\bea
a_{\bk j}\equiv \frac{1}{2\left|\eta_{\bk j}\right|^2}&\Big[&-\eta_{\bk j}\left|v_{\bk j}\right|^2(\left|\psi_{\bk j}\right|^2+\left|\psi_{-{\bk j}}\right|^2+1)-\no
&&(\left|\eta_{\bk j}\right|^2+u_{\bk j}\eta_{\bk j}v_{\bk j}^*)\psi_{\bk j}\psi_{-{\bk j}}\no 
&&+(\eta_{\bk j}^2-u_{\bk j}\eta_{\bk j}v_{\bk j})\psi_{\bk j}^*\psi_{-{\bk j}}^*\Big].\qquad\label{eq:prefactor_a}
\eea
Finally, we acquire the corresponding continuous form for $N\rightarrow\infty$, and the parameters then become continuous functions of time and we obtain the prefactor $a_{\bk j}\rightarrow a_{\bk}(t)$ of the squeezing part. Taking now the product over $\bk$ leads to the sum in the exponent and we arrive at the general two mode squeezed overlap of the Lagrangian
\bea\label{eq:overlap_continuum}
 \prod_{j=1}^{N}  \langle\Psi_j |\Psi_{j-1}\rangle
  & \rightarrow & \exp\Big[ \int_{t_{a}}^{t_{b}} dt \Big(\sum_{\bk \neq 0} \frac{\psi_{\bk} \partial_{t}\psi_{\bk}^{*} -\psi^{*}_{\bk} \partial_{t}\psi_{\bk}}{2}\nonumber\\
&&  + \sum_{\bk\neq 0}{\vphantom\sum}'(a_{\bk} \partial_{t}\eta_{\bk}^{*} - a^{*}_{\bk} \partial_{t}\eta_{\bk}) \Big)\Big]\, .
 \eea

\subsection{Expectation value $\expect{\Psi_j | H | \Psi_j}$}\label{appA2}
In the expectation value, succeeding times do not couple and we can omit the time index $j$ and consider our parameters to be continuous. Being in the weakly interacting regime at low temperature, we apply the Bogolyubov approximation to our Hamiltonian in Eq. ($2$) of the main text. The $\bk =0$ mode is approximated by the $c-$number $\sqrt{N_0}$ and the Hamiltonian is expanded up to second order in $b_{\bk}$ \cite{schwabl,xgw}. The Hamiltonian under consideration is then
\bea\label{eq:bogo_ham}
H=&&-\frac{gN_0^2}{2V}+\sum_{\bk\neq 0}(\epsilon_{\bk}+gn_0)b_{\bk}^{\dag}b_{\bk}\no
&&\,\,\quad+\frac{gn_0}{2}\sum_{\bk\neq 0}(b_{\bk}^{\dag}b_{-\bk}^{\dag}+b_{\bk}b_{-\bk}),
\eea
where $\epsilon_k = \hbar^2 k^2/(2m)$ is the free particle spectrum, $g$ is the interaction and $n_0$ is the condensate density.
The unitarity of the squeezing operator is manifest in the relation $S(\eta_\bk)^{\dag}=S(-\eta_\bk)$. Hence the transformation \eqref{eq:squeezetrafo} is modified to 
\beq\label{eq:squeezetrafo_modified}
S_{\bk}^{\dag}b_{\pm {\bk}}S_{\bk}=u_{\bk}b_{\pm {\bk}}+v_{\bk}b_{\mp {\bk}}^{\dag}. 
\eeq
With this, we can now calculate the squeezed Hamiltonian and employ the definition of the coherent state. The resulting expectation value reads as follows
\bea
&&\expect{\Psi |H|\Psi}= -\frac{gN_0^2}{2V}+\no
&&\sum_{k\neq 0}\left(\epsilon_k(u_k^2+\abs{v_k}^2)+gn_0\abs{u_k+v_k}^2\right)\abs{\psi_k}^2+\no
&&\left(\frac{gn_0}{2}(u_k^2+v_k^2)+(\epsilon_k+gn_0)u_kv_k\right)\psi_k^{*}\psi_{-k}^{*}+\no
&&\left(\frac{gn_0}{2}(u_k^2+v_k^{*2})+(\epsilon_k+gn_0)u_kv_k^*\right)\psi_k\psi_{-k}+\no
&&\epsilon_k\abs{v_k}^2+\frac{gn_0}{2}(\abs{u_k+v_k}^2-1).\label{eq:result_Ham}
\eea
This $c-$number valued Hamiltonian is diagonalized for the equilibrium value $\eta_{\bk}^{0}$ given in the main text. 

\subsection{Expansion around equilibrium}\label{appA3}
In the subsections \ref{appA1} and \ref{appA2}, we have calculated the overlap and the expectation value, respectively, and obtained a general two mode squeezed Lagrangian. For the weak interaction and low temperature regime, we analyze the system around the equilibrium parameter values up to second order. 

The prefactor $a_{\bk}$ in the overlap is expanded up to first order since it is multiplied by the time derivatives
\bea\label{eq:prefac_a_expanded}
a_{\bk\phantom{ ,1}}&=&a_{\bk ,1}\tilde{\eta}_k+\bar{a}_{\bk ,1}\tilde{\eta}_\bk^*+a_{\bk ,0}\, ,\no
&&a_{\bk ,0}=\frac{(\hbar\omega_\bk-\epsilon_\bk)^2}{8\hbar\omega_\bk\epsilon_\bk\eta_\bk^0}\no
&&a_{\bk ,1}=\frac{\hbar^2\omega_{\bk}^2-\epsilon_{\bk}^2}{8\hbar\omega_{\bk}\epsilon_{\bk}\eta_{\bk}^0}\\
&&\bar{a}_{\bk ,1}=a_{\bk ,1}-\frac{a_{\bk ,0}}{\eta_{\bk}^{0}}\, ,
\eea
where $\omega_{\bk}$ is the Bogolyubov dispersion.

We get the corresponding Hamiltonian by expanding Eq. \eqref{eq:result_Ham}
\bea\label{eq:expanded_Ham}
&&H\phantom{_0}\,\,\,=H_0+\sum_{\bk\neq 0}\hbar\omega_{\bk}\abs{\psi_{\bk}}^2+\sum_{\bk\neq 0}{\vphantom\sum}'(E_{\bk, r} \tilde{\eta}_{\bk, r}^{2} + E_{\bk, i} \tilde{\eta}_{\bk,i}^{2})\no
&&H_0\,\,\,=-\frac{gN_0^2}{2V}-\sum_{\bk\neq 0}\frac{(\hbar\omega_\bk-\epsilon_\bk)^2}{4\epsilon_\bk}\\
&&E_{\bk, r}=2\hbar\omega_{\bk}\\
&&E_{\bk, i}\,=\frac{(\hbar^2\omega_{\bk}^2-\epsilon_{\bk}^2)^2}{8\hbar\omega_{\bk}\epsilon_{\bk}^2(\eta_{\bk}^{0})^2}.
\eea
We can see clearly that the above Hamiltonian separates into two terms, one depending on solely the coherent field $H_{\psi}$ and another on the squeezing field $H_{\eta}$. Subtracting these from the corresponding overlap terms in Eq. \eqref{eq:overlap_continuum} give the Lagrangians in Eq. ($7$) and ($8$) of the main text. After applying the Euler-Lagrange equation, we obtain the two linear differential equations for the coherent parameter $\psi_{\bk}$ and for the squeezing parameter $\eta_{\bk}$. The dispersion of the coherent field can be read out directly. However, in case of the squeezing field, we transform the differential equation into its matrix form
\bea\label{eq:differentialequation}
&&\partial_t\begin{bmatrix}
\tilde{\eta}_{\bk,r} \\
\tilde{\eta}_{\bk,i}\\
\end{bmatrix}=\frac{1}{2\hbar}
\begin{bmatrix}
0 && \frac{E_{\bk, i}}{a_{\bk ,1}} \\
-\frac{E_{\bk, r}}{a_{\bk ,1}} && 0 \\
\end{bmatrix}\begin{bmatrix}
\tilde{\eta}_{\bk,r} \\
\tilde{\eta}_{\bk,i}\\
\end{bmatrix}.
\eea
Now, we can read out the dispersion immediately
\bea\label{eq:eta_dispersion}
\hbar^2 \omega_{\bk,\eta}^2 = \frac{E_{\bk,r} E_{\bk,i}}{4 a_{\bk, 1}^2}=4\hbar^2\omega_\bk^2.
\eea
This gives the dispersion $\hbar \omega_{\eta, \bk} = \sqrt{E_{\bk,r} E_{\bk,i}}/(2 a_{\bk, 1})=2\hbar\omega_\bk$, leading to the second sound velocity $c_2=2c_1$.

\subsection{Single-particle Green's function}\label{appA4}
In this section we show that the single-particle Green's function has poles associated with both sound modes. The single-particle Green's function is defined as
\bea\label{eq:single_particle_correlation_definition}
ig_1(\bk,t_2,t_1)\equiv\expect{0|T(b_\bk(t_2)b^\dag_\bk(t_1))|0}.
\eea
According to Ref. \cite{xgw} and \cite{mahan}, we can write the Green's function in terms of the evolution operator
\bea\label{eq:timeevolution}
ig_1(\bk,t_2,t_1)=\frac{\expect{0|T\left(b_\bk(t_2)b^\dag_\bk(t_1)U(\infty,-\infty)\right)|0}}{\expect{0|U(\infty,-\infty)|0}}\, .\qquad
\eea
This representation of $g_1$ enables us to construct the path integral in the squeezed coherent representation which renders the time expectation value an ensemble expectation value with classical fields
\bea\label{eq:timeevolution_classical}
ig_1(\bk,t_2,t_1)=\expect{\beta_\bk(t_2)\beta^*_\bk(t_1)}
\eea
with
\bea\label{eq:timeevolution_fields}
\beta_\bk(t)=u_\bk(t)\psi_\bk(t)+v_\bk(t)\psi_{-\bk}^*(t)\, .
\eea
The single-particle Green's function then reads as follows
\bea\label{eq:single_particle_correlation}
ig_1(\bk,t_2,t_1)&=&\expect{u_\bk(t_2)u_\bk(t_1)}\expect{\psi^*_\bk(t_1)\psi_\bk(t_2)}\no 
&+&\expect{v_\bk(t_2)v^*_\bk(t_1)}\expect{\psi_{-\bk}(t_1)\psi_{-\bk}^*(t_2)}\, ,\qquad
\eea
where we omit the $(\bk,-\bk)-$terms and write the squeezing and coherent correlations separately for reasons discussed in the next subsections.

\begin{figure*}[t]
\includegraphics[width=2.1\figurewidth]{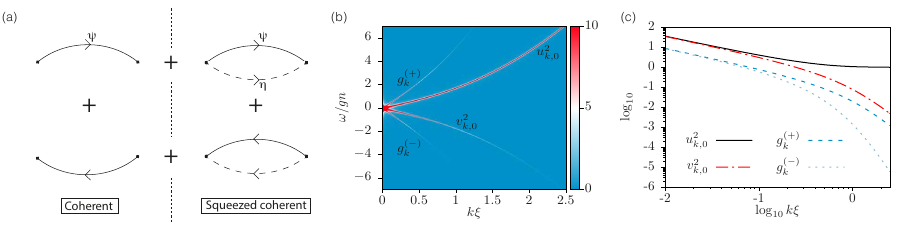}
\caption{(a) Illustration of Eq. \eqref{eq:final_greens_function_w} via Feynman diagrams, where the forward propagators correspond to particle and the backward to hole excitations. The inclusion of the squeezing field complements the "coherent propagator" (continuous lines) with the "squeezing propagator" (dashed lines). (b) $\abs{g_1(k\xi,\omega)}$ as a function of $\omega$ and dimensionless $k\xi$. The four branches correspond to the poles of $g_1(\bk,\omega)$. For increasing $\bk$, the branches that correspond to the poles $\pm(\omega_\bk+\omega_{\bk,\eta})$ vanish first, followed by the pole $-\omega_\bk$. The branch that corresponds to the pole $+\omega_\bk$ remains. (c) Weights of the branches in (b) as a function of $k\xi$. The weights $g_\bk^{(\pm)}$ that emerged from the inclusion of the squeezing field decay very quick, followed by $v_{\bk,0}^2$. The only surviving weight, i.e., $u_{\bk,0}^2\rightarrow 1$, gives rise to the single-particle Green's function of a free particle.
}
\label{fig:green}
\end{figure*}

\subsubsection{Expectation value}
In order to evaluate the above single-particle Green's function, we write its corresponding expectation value in the frequency domain as follows
 \bea\label{eq:expect}
 &\expect{\chi^*_{\bk_1,\omega_1}\chi_{\bk_2,\omega_2}}&=\frac{\int \mathcal{D}^2(\psi,\eta)\, e^{\frac{i}{\hbar}\int_{-T/2}^{T/2}dt\, \mathcal{L}}\chi^*_{\bk_1,\omega_1}\chi_{\bk_2,\omega_2}}{Z}\, \no
 &\text{with   }\mathcal{D}^2(\psi,\eta)&\equiv\prod_{k_x>0,\omega} d^2\psi_{\bk,\omega}d^2\psi_{-\bk,\omega}d^2\eta_{\bk,\omega}\, ,\qquad
 \eea
 following Ref.  \cite{altlandsimons}. And the corresponding partition function is
  \bea\label{eq:partition_function}
 Z=\int\mathcal{D}^2(\psi,\eta)\,e^{\frac{i}{\hbar}\int_{-T/2}^{T/2}dt\, \mathcal{L}}\, ,
 \eea 
 where the complex fields $\chi_{\bk,\omega}$ represent the coherent $\psi_{\bk,\omega}$ and squeezing fields $\eta_{\bk,\omega}$ and $\mathcal{L}=\mathcal{L}_\psi+\mathcal{L}_\eta$ is the expanded Lagrangian around equilibrium, mentioned in the main text. Our calculation simplifies, when $\mathcal{L}$ is diagonal. Therefore, we write our fields in their Fourier series representation
 \bea\label{eq:fields_fourier_series}
\chi_\bk (t)=\frac{1}{\sqrt{T}}\sum_{n} \chi_{\bk ,\omega_n}e^{-i\omega_n t}\, ,
\eea
with $\omega_n=2\pi n/T$. If the fundamental period is $T\rightarrow\infty$, then the frequency steps $2\pi/T$ become infinitesimal and thus the fields are continuous in $\omega$. Upon integrating over the fundamental period  $T$, only the coherent part is already diagonal
\bea\label{eq:lagrangian_psi_action}
S_\psi\equiv\int_{-T/2}^{T/2} dt\, \mathcal{L}_\psi =\hbar\sum_{\bk \neq 0,\omega} (\omega - \omega_{\bk}) \psi^*_{\bk ,\omega} \psi_{\bk ,\omega}\, .\qquad
\eea
For the squeezing part, we introduce the following transformations of the real and imaginary parts
\bea\label{eq:new_squeezing_fields}
\tilde{\eta}_{\bk,r}&=&\left(\frac{E_{\bk,i}}{4E_{\bk,r}}\right)^{\frac{1}{4}}\left(\xi_{\bk}+\xi_{\bk}^*\right)\no
\tilde{\eta}_{\bk,i}&=&-i \left(\frac{E_{\bk,r}}{4E_{\bk,i}}\right)^{\frac{1}{4}}\left(\xi_{\bk}-\xi_{\bk}^*\right).
\eea
After inserting these into $\mathcal{L}_\eta$ and then integrating, we obtain
\bea\label{eq:lagrangian_eta_action}
S_\eta\equiv \sum_{\bk\neq 0,\omega}{\vphantom{\sum}}'4a_{\bk ,1}\left(\hbar\omega-\frac{\sqrt{E_{\bk,r}E_{\bk,i}}}{2a_{\bk ,1}}\right)\xi_{\bk,\omega}^*\xi_{\bk,\omega}.\qquad
\eea
Hence, the expression in the parenthesis vanishes for the same dispersion derived in Eq. \eqref{eq:eta_dispersion}
\bea\label{eq:pole}
\omega=\frac{\sqrt{E_{\bk,r}E_{\bk,i}}}{2a_{\bk ,1}\hbar}\, .
\eea

\subsubsection{Wick's theorem for squeezed coherent fields}

\begin{figure*}[t]
\includegraphics[width=2.1\figurewidth]{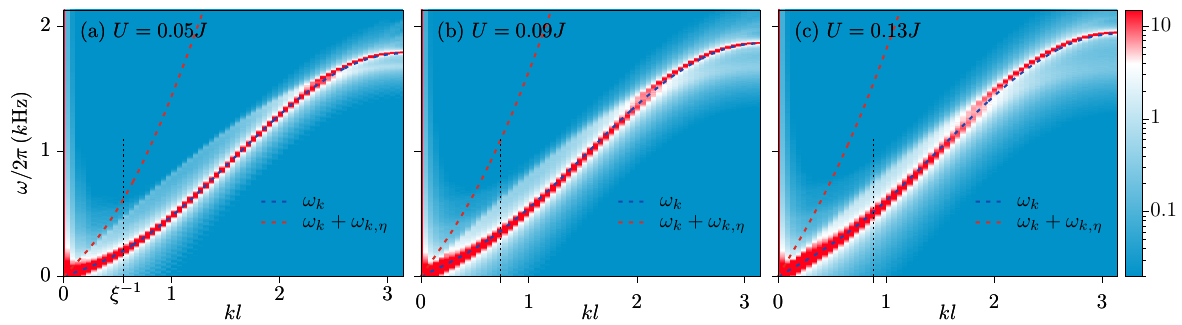}
\caption{ Single-particle correlation function $c_{1}(\bk, \omega)$ as a function of frequency $\omega$ and momentum $k_{x}$, for the interactions $U/J=0.05$, $0.09$, and $0.13$. The two excitation branches are compared to the Bogolyubov dispersion $\omega_{\bk}$, lower (blue) dashed line, and the side peak at $\omega_{\bk} + \omega_{\bk, \eta}$, upper (red) dashed line, motivated by the pole structure of Eq. \eqref{eq:final_greens_function_w}. 
The vertical dashed lines denote the inverse healing length $\xi^{-1}$.
}
\label{fig:g1_num}
\end{figure*}

We evaluate expectation values of the form \eqref{eq:expect} by calculating the expectation value of the source terms $S_\rho$ by using Eq. \eqref{eq:gaussianrelation}. The result is
\bea\label{eq:sourceterms}
\expect{e^{S_\rho}}&=&e^{i\sum_{k_x>0,\omega}\frac{\bar{j}_{\bk,\omega}j_{\bk,\omega}}{(\omega-\omega_{\bk}+i\delta)}+i\sum_{\bk,\omega}'\frac{\bar{\ell}_{\bk,\omega}\ell_{\bk,\omega}}{4a_{\bk ,1}(\omega-\omega_{\bk,\eta}+i\delta)}}\no
\text{with}\no
S_\rho &\equiv &\sum_{\bk\neq 0,\omega}\bar{j}_{\bk,\omega}\psi_{\bk,\omega}+j_{\bk,\omega}\psi^*_{\bk,\omega}\no 
&&+\sum_{\bk,\omega}{\vphantom{\sum}}'\bar{\ell}_{\bk,\omega}\xi_{\bk,\omega}+\ell_{\bk,\omega}\xi^*_{\bk,\omega} \, .
\eea
With this relation we can write the Wick formula for the expectation values of the coherent and squeezing fields
\bea\label{eq:moments}
&&\langle\big(\psi^*_{\bk_1,\omega_1}\big)^m\big(\psi_{\bk_2,\omega_2}\big)^n\big(\xi^*_{\bk_3,\omega_3}\big)^r\big(\xi_{\bk_4,\omega_4}\big)^w\rangle\no 
&&=\left[\partial_{\bar{\ell}_{\bk_4,\omega_4}}^{(w)}\partial_{\ell_{\bk_3,\omega_3}}^{(r)}\partial_{\bar{j}_{\bk_2,\omega_2}}^{(n)}\partial_{j_{\bk_1,\omega_1}}^{(m)}\expect{e^{S_\rho}}\right]_{\substack{\bar{j}_{\bk_2,\omega_2}=0 \\ j_{\bk_1,\omega_1}=0 \\ \bar{\ell}_{\bk_4,\omega_4}=0 \\ \ell_{\bk_3,\omega_3}=0}}\, ,\qquad
\eea
where we shifted the roots of $S_\psi$ and $S_\eta$ into the complex plane by adding the damping term $\delta$. Therefore, we have
\bea\label{eq:expectationvalues}
\langle\psi_{\bk_1,r\omega_1}^*\psi_{\bk_2,\omega_2}\rangle&=&\frac{i\delta_{\bk_1,\bk_2}\delta_{\omega_1,\omega_2}}{(\omega_1-\omega_{\bk_1}+i\delta)}\no
\langle\xi_{\bk_1,\omega_1}^*\xi_{\bk_2,\omega_2}\rangle&=&\frac{i\delta_{\bk_1,\bk_2}\delta_{\omega_1,\omega_2}}{4a_{\bk,1}(\omega_1-\omega_{\bk_1,\eta}+i\delta)}\, .\qquad
\eea
According to our formula in Eq. \eqref{eq:moments} the expectation value for $m=n=r=w=1$ breaks down into a product of squeezing and coherent fields. This was to be expected because the Lagrangian does not include the coupling between the fields. The expressions in Eq. \eqref{eq:expectationvalues} are related with the original expectation values in Eq. \eqref{eq:single_particle_correlation} via Fourier transformations of the type
\begin{align}\label{eq:expectationvalues_time}
&\expect{\chi^*_\bk(t_1)\chi_\bk(t_2)}\no 
&=\frac{1}{T}\sum_{\omega_n,\omega_m}\expect{\chi^*_{\bk,\omega_n}\chi_{\bk,\omega_m}}e^{i\omega_n t_1-i\omega_m t_2}\, .\qquad
\end{align}
Knowing the Fourier representation of the ensemble expectation value, we switch now into time domain and obtain
\bea\label{eq:expectationvalues_time}
\expect{\psi^*_\bk(t_1)\psi_\bk(t_2)}&=&\Theta(t_2-t_1)\, e^{-i\omega_\bk (t_2-t_1)}\no 
\expect{\psi_\bk(t_1)\psi^*_\bk(t_2)}&=&\Theta(t_1-t_2)\, e^{-i\omega_\bk (t_1-t_2)}\no 
\expect{\xi^*_\bk(t_1)\xi_\bk(t_2)}&=&\Theta(t_2-t_1)\, \frac{e^{-i\omega_{\bk,\eta} (t_2-t_1)}}{4a_{\bk,1}}\no 
\expect{\xi_\bk(t_1)\xi^*_\bk(t_2)}&=&\Theta(t_1-t_2)\, \frac{e^{-i\omega_{\bk,\eta} (t_1-t_2)}}{4a_{\bk,1}}\, .
\eea
Since the Green's function is time translational invariant, we write from here on $t\equiv t_2-t_1$.
Next, similar to the Lagrangian, we expand the Green's function in Eq. \eqref{eq:single_particle_correlation} to second order and after inserting the above relations, the single-particle Green's function in the time domain is
\bea\label{eq:single-particle_green_time}
&ig_1(\bk,t)=\Theta(t)\Big(u_{\bk,0}^2e^{-i\omega_\bk t}+g_\bk^{(+)}e^{-i\omega^+_\bk) t}\Big)&\no 
&+\Theta(-t)\Big(v_{\bk,0}^2e^{i\omega_\bk t}+g_\bk^{(-)}e^{i\omega^+_\bk t}\Big)\, ,&\qquad
\eea
where the newly emerged dispersions are $\omega^\pm_\bk\equiv\omega_\bk\pm\omega_{\bk,\eta}$. In Eq. \eqref{eq:single-particle_green_time}, we omitted the $\Theta(t)\Theta(-t)$ terms as they vanish in the Fourier transform. Consequently, the dispersion $\omega^-_\bk$ vanishes. The prefactors are
\bea\label{eq:prefacs}
g_\bk^{(+)}&=&\frac{v_{\bk,0}^2}{4}\qquad g_\bk^{(-)}=\frac{v_{\bk,0}^4}{4u_{\bk,0}^2}\no
u_{\bk,0}&=&\frac{\hbar\omega_\bk+\epsilon_\bk}{2\sqrt{\hbar\omega_\bk\epsilon_\bk}}\qquad v_{\bk,0}=\frac{\hbar\omega_\bk-\epsilon_\bk}{2\sqrt{\hbar\omega_\bk\epsilon_\bk}}\, .
\eea
We can now extract the poles by considering the Fourier transform of the Green's function
\bea\label{eq:Green_ft}
g_1(\bk,\omega)=\int_{-\infty}^\infty dt e^{i\omega t-\delta\abs{t}}\, g_1(\bk,t)\, ,
\eea
where we introduced the small damping parameter $\delta>0$. Consequently, the single-particle Green's function is
\bea\label{eq:final_greens_function_w}
g_1(\bk,\omega)&=&\underbrace{\frac{u_{\bk,0}^2}{\omega-\omega_\bk+i\delta}-\frac{v_{\bk,0}^2}{\omega+\omega_\bk-i\delta}}_{=g^B_1(\bk,\omega)}\no 
&+&\frac{g_\bk^{(+)}}{\omega-\omega^+_\bk+i\delta}-\frac{g_\bk^{(-)}}{\omega+\omega^+_\bk-i\delta}\, .
\eea
The first two terms correspond to the single-particle Green's function of the Bogolyubov case which is the usual result in the coherent state representation. However, after including the squeezing field into the path integral, we get an additional Green's function that has poles at the sum of the Bogolyubov and the squeezing dispersions. In Fig. \ref{fig:green}(a), we illustrate Eq. \eqref{eq:final_greens_function_w} with Feynman diagrams. The inclusion of the squeezing fields corresponds to considering higher order diagrams, extending the Bogolyubov picture. In (b) we depict the absolute value of $g_1(\bk,\omega)$, where we have set the free particle dispersion to a dimensionless value $\epsilon_\bk\rightarrow \epsilon_\bk/gn=k^2\xi^2$. The two branches in the middle of figure (b) are the Bogolyubov peaks at the poles of $g_1^B(\bk,\omega)$ in Eq. \eqref{eq:final_greens_function_w} and the side peaks correspond to the poles of the newly emerged terms. For large $\bk$, the side peaks decay. Upon increasing $\bk$ further, the lower Bogolyubov peak falls off and the upper Bogolyubov peak survives and coincides with the free particle Green's function. In panel (c), the corresponding weights are shown for comparison.

\section{Simulated single-particle correlation function}

In this section, we determine the single-particle correlation function $c_{1}(\bk, \omega)$ using the simulation technique described in the main text, and discuss its dependence on the interaction strength $U/J$.
The single-particle correlation function $c_{1}(\bk, \omega)$ is defined in the main text. We consider a homogeneous condensate of $^{7}$Li atoms with density $\rho = 0.6\times 10^{13}$cm$^{-3}$ and temperature $T/J = 4.5$, which are the same as in the main text.
We determine $c_{1}(\bk, \omega)$ at the three different interaction strengths $U/J= 0.05$, $0.09$, and $0.13$, and show these results in Fig. \ref{fig:g1_num}. 
We observe two excitation branches: the Bogolyubov and second sound mode, which are compared to the Bogolyubov dispersion $\omega_{\bk}$, and the side band $\omega_{\bk}^+$ that was found in Eq. (\ref{eq:final_greens_function_w}). 
The result in panel (a) corresponds to $U/J= 0.05$ and is the same as the main text, whereas panels (b) and (c) correspond to $U/J= 0.09$ and  $0.13$, respectively.  
In the low-energy regime, the analytical second sound dispersion agrees for $U/J= 0.05$, whereas it deviates systematically for high interactions due to higher order terms of the Lagrangian, which are not included in this weak-coupling limit of the dispersions. 
Furthermore, in the high-energy regime, we observe in the numerical $c_{1}(\bk, \omega)$ a crossing between the Bogolyubov and second sound mode as a function of wavevector $k_x$. 
Beyond this crossing, the second sound mode occurs below the Bogolyubov mode, which is an indication of the onset of the hydrodynamic regime where second sound is below first sound.
This crossing is shifted to low $k_x$ values for high interactions, 
which suggests that for strong interactions one would recover the standard hydrodynamic low temperature result $c_2/c_1 = 1/\sqrt{3}$, as in the case of liquid helium.



\begin{thebibliography}{99}
\bibitem{footnote}
It can be realized with $\sideset{}{^{\prime}}\prod_{\bk\neq 0} =\prod_{\bk\neq 0, k_{x}>0}$, or any other way of restricting the product to half of the momentum states.
\end{thebibliography}
\end{document}